\documentclass[aps,pra,twocolumn,showpacs,superscriptaddress,groupedaddress]{revtex4}
\usepackage{graphicx}  
\usepackage{dcolumn}   
\usepackage{bm}        
\usepackage{amssymb}   
\usepackage{amsmath}
\usepackage{bm}
\usepackage{bbm}

\newcommand\A{\ensuremath{\bm{A}}}

\newcommand\n{\ensuremath{\mathbf{n}}}

\newcommand\bn{\ensuremath{\bm{\nabla}}}

\renewcommand\i{\ensuremath{\textup{i}}}

\newcommand\dfn{\ensuremath{\mathrel{\mathop:}=}}

\newcommand\J{\ensuremath{\mathcal{J}}}
\newcommand\K{\ensuremath{\mathcal{K}}}
\newcommand\PP{\ensuremath{\mathcal{P}}}
\newcommand\conj[1]{\ensuremath{\overline{#1}}}

\begin{document}
\title{Exploration of stable and unstable vortex patterns in a
  superconductor under a magnetic disc}

\author{Nico Schl\"omer}
\affiliation{Dept. Mathematics and Computer Sciences, Universiteit
  Antwerpen, 2020 Antwerpen, Belgium} 
\author{Milorad~V.~Milo\v{s}evi\'c
}\affiliation{Dept. Physics, Universiteit Antwerpen, 2020 Antwerpen,
  Belgium}
\author{Bart Partoens} \affiliation{Dept. Physics,
  Universiteit Antwerpen, 2020 Antwerpen, Belgium} 
\author{Wim Vanroose} \affiliation{Dept. Mathematics and Computer Sciences,
  Universiteit Antwerpen, 2020 Antwerpen, Belgium} \date{\today}

\begin{abstract}
  The stable and unstable solutions of a square 2D extreme type-II
  superconductor are studied in the field of a magnetic disc.  We use
  a preconditioned Newton-Krylov solver to find the solutions and use
  numerical continuation to track the solutions as the field strength
  varies.  For a disc with a small radius, we have identified generic
  scenarios through which the state loses and regains its stability.
\end{abstract}

\maketitle

\section{Introduction}
\label{sec:intro} Mesoscopic phenomena have been at the forefront of
research in superconductivity in the last 15
years.\cite{grigorieva1997phase,moschalkov1995effect,schweigert1998vortex,baelus2002dependence,kanda2004experimental,mel2002mesoscopic}
The key property of mesoscopic samples is the strong influence of
confinement on the superconducting condensate, and subsequent quantum
tailoring of superconductivity. There it is particularly important
that critical parameters can be enhanced by confinement. For example,
critical magnetic field of mesoscopic superconductors can be multiply
larger than one of bulk materials. In addition, since the size of a
mesoscopic sample is comparable to the scale of the vortex-vortex
interaction, there is a strong influence of the boundary on the flux
distribution in the sample (i.e. vortex configurations). Additional
influence on the aforementioned phenomena can be made by
nanoengineering, e.g. by introducing holes in the
sample~\cite{baelus2000vortex,xu2008magnetic}, or magnets embedded in,
or placed on top of the superconductor~\cite{milosevic2007stabilized}.
Namely, the strongly inhomogeneous magnetic field of the ferromagnet
interacts with externally applied magnetic field and changes the
magnetic environment for the superconductor, while also altering the
spatial domain for emerging vortices and stimulating nucleation of
vortex-antivortex pairs.

The subject of superconductor-ferromagnet hybrids is very mature,
and several reviews are available \cite{velez2008superconducting,lyuksyutov2005ferromagnet,aladyshkin2009nucleation}. However, the
literature to date has studied mainly the stable vortex(-antivortex)
configurations and how they depend on the parameters of the system
such as field strength or geometry. Virtually nothing is known about
how these states behave when they lose their stability and how the
transitions between stable states occur, i.e. how vortices nucleate
or enter the system, and how the energy barriers for those processes
can be quantified. Understanding this is clearly important for any
emerging devices that control or exploit the behavior of the
vortices \cite{milosevic2010vortex}. 

To arrive at a complete picture it is thus necessary to know the
location and evolution of the unstable or so-called saddle-point
states. Such states and their free energies directly determine the
possible transitions and the height of the energy barrier that
separate the stable states. An approach to study the saddle points
has been proposed in Ref. \cite{schweigert1999flux} and utilized in Ref.
\cite{baelus2001saddle}, but only for 2D superconducting disks. The radial
symmetry of the sample was crucial for the efficiency of the
approach, although theoretical formalism is generally applicable
once the eigenfuctions of the linearized Ginzburg-Landau equation
are known for the chosen geometry.

One more reason to study the saddle points in detail is the fact
that knowledge about the unstable states and their bifurcations
together with the symmetries of the system allows to analyze
superconductors by tools of the mathematics of dynamical systems.
One can then use the known techniques from pattern
formation,\cite{Hoyle:2006:PF} which would further expand the
scientific community and cross-fertilization of different subjects
and fields of study.

Therefore, in this paper we demonstrate the use of a new
high-performance solver for both stable and unstable solutions of
the Ginzburg-Landau equation, for an arbitrary superconducting
geometry, and in presence of an arbitrary magnetic field. As an
exemplary case, we choose superconducting square with a magnetic
disc on top, with additional possibility of added homogeneous
magnetic field. Important part of the presented study is the method
of numerical continuation of relevant parameters, which has been
successfully used for Bose-Einstein condensates \cite{law2010stable,PhysRevA.68.023609}. The solver we use is proposed in \cite{SV:2012:OLS} 
and is available as a open-source library at \cite{nosh}

The outline of the paper is follows. In Sec. II we describe the
system, its numerical representation and how the resulting system
can be solved using a Newton-Krylov iteration. We then show how to
track the solutions as the parameters of the system change in
Sec.~\ref{sec:numericalcontinuation}. In Sec. III we show the stable
and unstable solutions, and transitions between them, for a square
superconducting domain in external homogeneous field. The results
for a hybrid structure are given further, for a small magnetic disc
in Sec.~\ref{sec:radius2.0} and for a large disc in
Sec.~\ref{sec:radius4.5}. We conclude with a discussion and summary
in Sec. VI.

\begin{figure}
\includegraphics[width=0.5\textwidth]{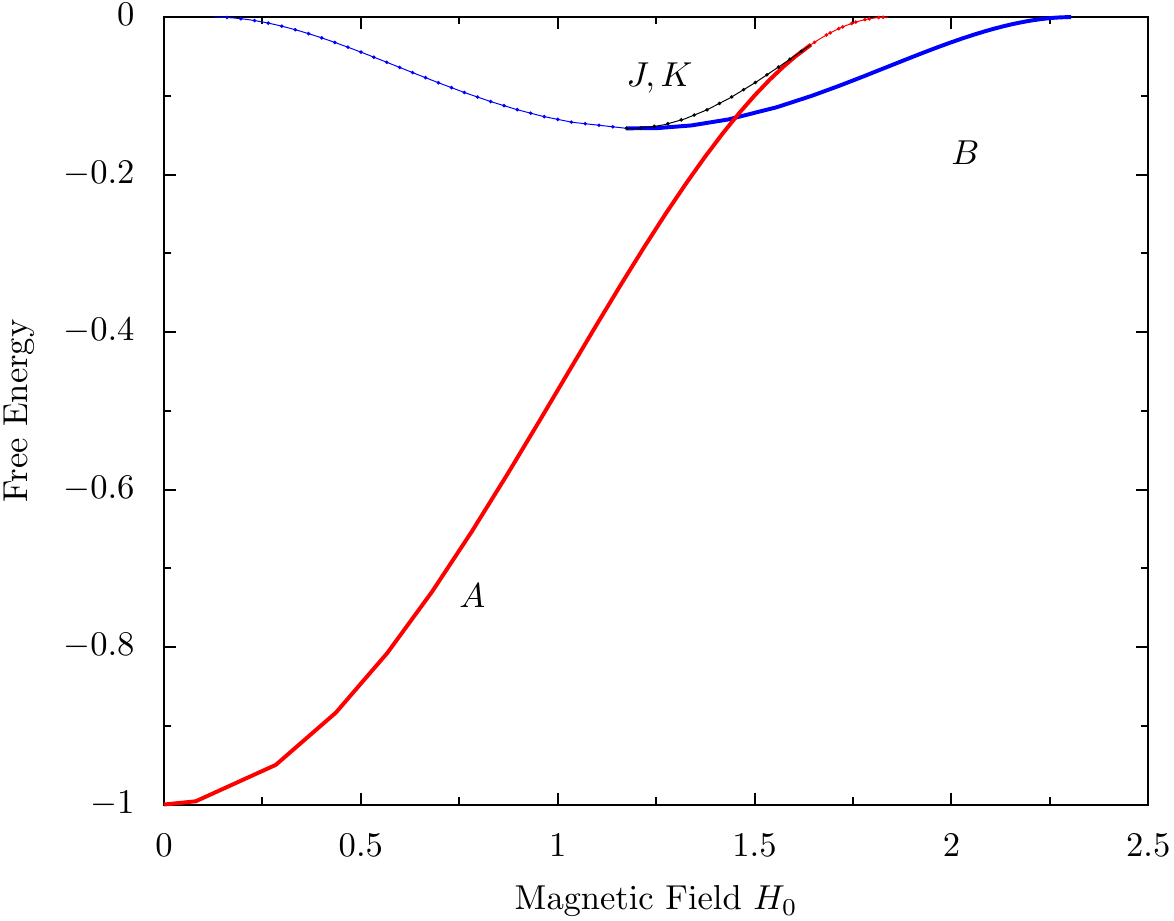}\\
\includegraphics[width=0.5\textwidth]{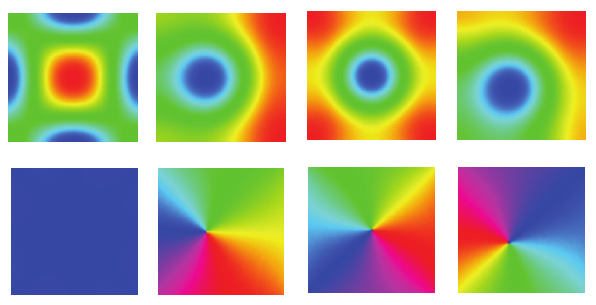}
\caption{(Color online) The stable and unstable states in a $3\xi \times 3\xi$
  superconducting square sample in a homogeneous field. The branch $A$
  loses it stability for large fields, while the branch $B$ that has a
  single vortex in the center of the domain gains stability for large
  fields. The two bifurcation points are connected through two unstable
  branches $J$ and $K$. The corresponding patterns are shown below
  where we show from left to right the pattern for curve $A$, $J$ and
  $B$ and $K$ respectively.  Curve $J$ has a pattern entering along
  the center line, while curve $K$ has it entering along the
  diagonal. We refer to \cite{SAV:2012:NBS} for more details.}
\label{fig:small}
\end{figure}
\section{The Ginzburg-Landau equation and its numerical solution}
\subsection{Description of the system}
We look at extreme type II superconductors where the magnetic field is
not modified by the super-currents that are generated by the
superconductor.  In this case, the Ginzburg--Landau problem decouples
and results in
\begin{equation}\label{eq:GL}                                                             
  \mathcal{GL}(\psi) = 
\begin{cases}                                                                             
  0 = \left(-\i\bn - \A\right)^2 \psi - \psi \left(1 - |\psi|^2\right)
  \quad \text{in }
  \Omega,  \\[3mm]
  0 = \n \cdot ( -\i\bn - \A) \psi \quad \text{on
  } \partial\Omega, \end{cases} \end{equation} where $\A$ is the given
magnetic vector potential and $\psi$ is the order parameter
describing the superconducting state.

In this paper we study problems where the magnetic field is
generated by a magnetic disc.  The vector potential is then an
integral of the magnetic dipole over the volume $V$  of the disc
\begin{equation}
  \A(\mathbf{x}) = \int_V \frac{\mathbf{m}\times (\mathbf{x}-\mathbf{z})}{\|\mathbf{x}-\mathbf{z}\|^{3/2}} d\mathbf{z}
\end{equation}
where $\textbf{m}$ is the magnetic moment. In this example it is
oriented along the $z$-axis.

The magnetic disc has a height $h$ and a radius $r$ and sits a distance
$z$ above the center of the square domain.

\subsection{Newton--Krylov method}
\label{sec:newton-krylov}
We use a Newton-Krylov iterative method to find the solution of
\eqref{eq:GL}.  The Newton iteration is an outer iteration that deals
with the nonlinearity of the equation where each iteration requires
the solution of the Jacobian system.  The Krylov subspace method is an
inner iteration where the Jacobian system is solved iteratively. 
The advantage of this method is that the Newton method converges
superlinearly in the neighborhood of a steady state, irrespective of
the stability properties of the solution. It starts from an initial
guess $\psi^{(0)}$ and a sequence of approximations $\psi^{(k+1)} =
\psi^{(k)} + \delta \psi^{(k)}$ is generated. The update $\delta
\psi^{(k)}$ satisfies the linear system
\begin{equation}\label{eq:newton}
\J(\psi^{(k)}) \delta \psi^{(k)} = -\mathcal{GL}(\psi^{(k)}).
\end{equation}
The Jacobian $\J$ is a linear operator defined via the action
\begin{equation}\label{eq:schroed J}
\J(\psi)\phi = (\K + V + 2g|\psi|^2) \phi + g \psi^2 \overline{\phi},
\end{equation}
where $\overline{\phi}$ denotes complex conjugation.

Solving the large linear system (\ref{eq:newton}) in a scalable way is
the most significant difficulty when applying Newton's method to
nonlinear Schr\"odinger equations. However, the linear systems such
as~(\ref{eq:newton}) can be solved using Krylov iterative methods that
do not require an explicit (matrix) representation of the operator but
only its application to vectors such as~(\ref{eq:schroed J}).  The
convergence of those methods is highly dependent on the spectrum of
the involved linear operator.

To solve the Jacobian system, we use the method proposed in
\cite{SV:2012:OLS} where it is shown that $\J$ from
Eq.~(\ref{eq:schroed J}) associated with the nonlinear Schr\"odinger
equations is self-adjoint with respect to the inner product
\begin{equation}\label{eq:scalar product}
\langle \phi, \psi\rangle \dfn \Re \left( \int_\Omega \conj{\phi} \psi \right).
\end{equation}
The paper suggests to use MINRES \cite{greenbaum1997iterative}, a
Krylov subspace method suitable for indefinite self-adjoint
problems. 

One characteristic of Krylov methods is that a larger number of
unknowns increases the number of iterations that are needed to achieve
convergence.  In addition, the computational cost of a single
evaluation of the linear action also grows with the number of
unknowns. Therefore, high-resolution discretizations of
three-dimensional systems would require a significant computational
effort. 

In \cite{SV:2012:OLS} a \emph{preconditioner} is proposed for the
linear problem.  The main idea is that, instead of solving the
discretized version $J x = b$ of (\ref{eq:newton}), one can solve an
equivalent, numerically more favorable problem $P^{-1}J x = P^{-1}b$
with a linear, invertible preconditioning operator $P$.  If $P$ is
appropriately chosen, Krylov methods applied to the new system
converge much faster. In the case of the linearization of nonlinear
Schr\"odinger operators (\ref{eq:schroed J}), the energy operator $\K$
is of particular interest, as it typically dominates the spectral
behavior of $\J(\psi)$. More precisely, define the symmetric
preconditioning operator
\begin{equation}\label{eq:prec}
\PP(\psi) \dfn \K + 2\max\{g,\varepsilon\}|\psi|^2,
\end{equation}
with $0<\varepsilon\ll 1$ \cite{SV:2012:OLS}. Note that $\PP(\psi)$ is
strictly positive-definite except for the uninteresting case of
$\psi\equiv 0$. This, most importantly, makes the inversion of the
discretized $\PP(\psi)$, $P^{-1}(\psi)$, computationally cheap since
its positive-definiteness makes it a suitable target for geometric or
algebraic multigrid (AMG) solvers that yield optimal convergence
\cite{trottenberg2001multigrid}. As will be shown, even inexact
inversions of~(\ref{eq:prec}) used as preconditioners
for~(\ref{eq:schroed J}) make the Krylov convergence \emph{independent
  of the number of unknowns.}

\begin{figure}
\includegraphics[width=0.5\textwidth]{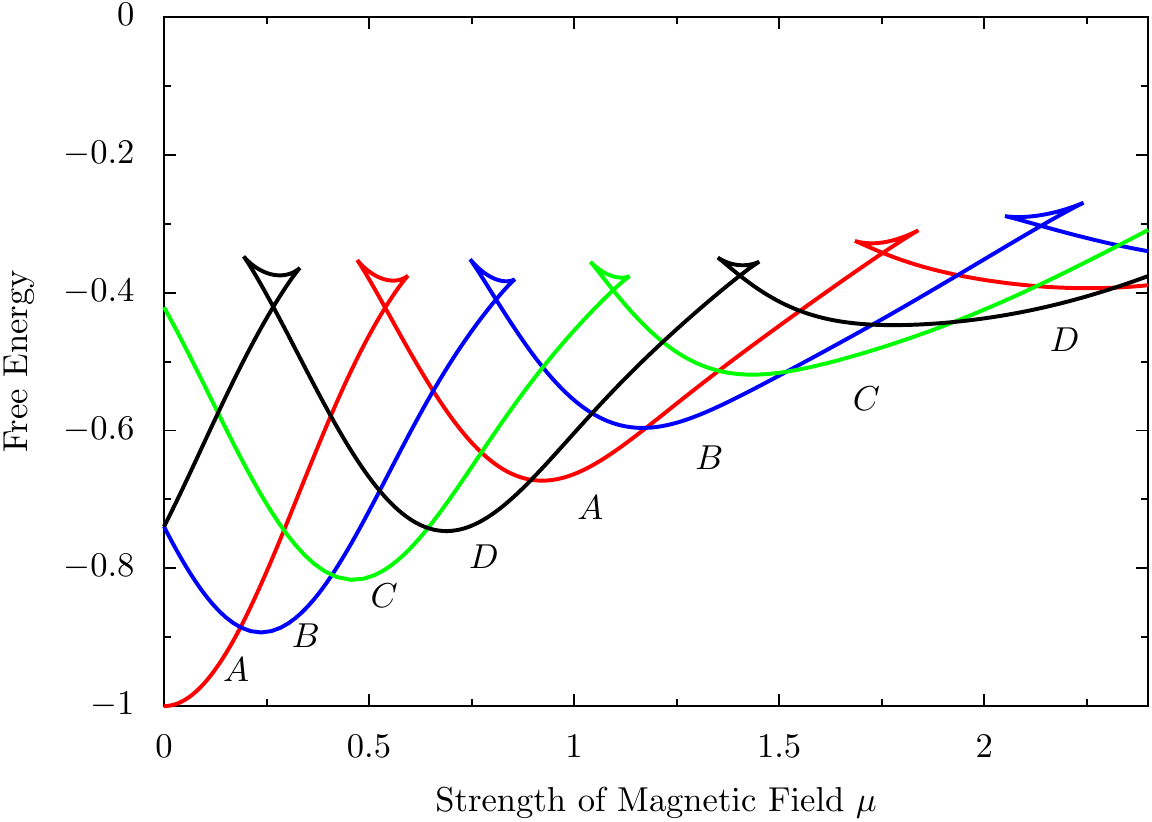}
\includegraphics[width=0.5\textwidth]{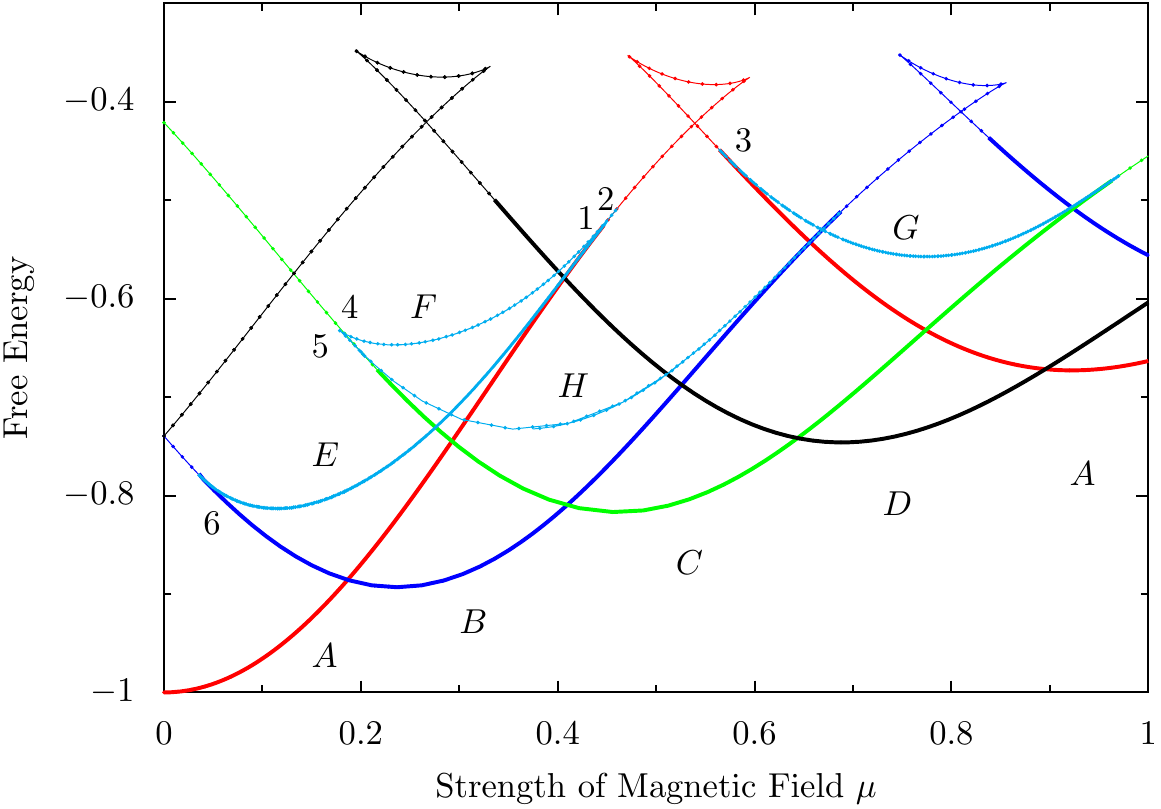}
\caption{(Color online) Top: over view of the states of a $10\xi
  \times 10\xi$ superconducting sample under a magnetic disc of radius
  $r=2\xi$.  The curve $A$, $B$, $C$ and $D$ have full symmetry of the
  system and they form swallow tails where stability is lost and
  regained.  Bottom: Detail for weak fields. Solid curves are stable
  and dotted curves are unstable.  Representative patterns along each
  of these curves are shown in Fig.~\ref{fig:small_dot}.  Between the
  bifurcation points, labeled as 1, 2, 3, \ldots, there are connecting
  curves with patterns with a reduced symmetry.  The curves $E$, $F$,
  $G$ and $H$ are unstable and have an additional vortex of multiple
  vortices moving in along one of the center lines.  Representative
  patterns for these curves are shown in
  Fig.~\ref{fig:small_connecting}}
 \label{fig:rad2}
\end{figure}
\subsection{Numerical Continuation}
\label{sec:numericalcontinuation}
The efficient linear solver from \cite{SV:2012:OLS} can be combined
with numerical parameter continuation into an efficient method for the
exploration of the energy landscape. Numerical parameter continuation
is a well-established technique for numerical bifurcation analysis of
dynamical systems \cite{keller1987lectures, Krauskopf2007}. It extends
the non-linear system to include parameters like the magnetic field
strength.  A function $F: \mathbb{C}^{N} \times \mathbb{R} \rightarrow
\mathbb{C}^N$ is considered that maps $\mathbf{x} := (\psi, \mu)$, a combination of
the numerical representation of $\psi$ and the parameter $\mu$ of the
magnetic field, to zero:
\begin{equation}
  F(\mathbf{x}) = 0
\end{equation}
By the implicit function theorem, this defines a family of solutions
$\psi(\mu)$ that can be generated numerically.

Numerical continuation follows a predictor-corrector scheme to
construct, starting from an initial solution point
$\mathbf{x}_0=(\psi_0,\mu_0)$, the successive points on the solution
curve.  The predictor step is an Euler predictor that uses the unit
length tangent vector $\mathbf{t}_i$ to the curve at a solution point
$\mathbf{x}_i$ that satisfies $F(\mathbf{x}_i) = 0$ and a step size
$s > 0$ to predict a guess $\mathbf{x}^\text{P}_{i+1}$ for the next point
on the curve
\begin{equation}
  \mathbf{x}_{i+1}^\text{P} = \mathbf{x}_i +  s \mathbf{t}_ i .
\end{equation}
The corrector step improves the guess $\mathbf{x}^\text{P}_{i+1}$ with
a Newton iteration on the augmented system to obtain a new solution
point $\mathbf{x}_{i+1}$. This augmented system has, in addition to
the constraint that $F(\mathbf{x}_{i+1}) = 0$, the requirement that
$\mathbf{x}_{i+1}$ must lie on the hyper-plane through
$\mathbf{x}^\text{P}_{i+1}$ perpendicular to $\mathbf{t}_i$ , the
tangent to the previous solution. This translates into the system
\begin{equation}
\begin{cases}\label{eq:numerical_continuation}
 F(\mathbf{x}_{i+1}) = 0\\
 (\mathbf{x}_{i+1} - \mathbf{x}^\text{P}_{i+1})^\text{T}\cdot \mathbf{t}_ i = 0,
\end{cases}
\end{equation}
where $\cdot$ is the dot-product and $T$ denotes the transpose. This
system is a map from $\mathbb{C}^n \times \mathbb{R}$ to $\mathbb{C}^n
\times \mathbb{R}$ and defines, under some conditions that are usually
met, uniquely the next point on the solution curve
\cite{Krauskopf2007}.

The system is solved again with a Newton-Krylov procedure. The
Jacobians of this augmented system is very similar to the Jacobian of
the Ginzburg-Landau equation. It is only extended with one row and column.  The
column contains the derivatives of the system to the parameters. The
additional row depends on the tangent that appears in the
system~\eqref{eq:numerical_continuation}.  Again the system can be
solved with a Newton-Krylov iteration.

Repeated application of a continuation step makes it possible to
follow a solution of the Ginzburg-Landau equation as the strength of
the magnetic field changes.  Since Newton iterations are insensitive
to the states' stability properties we are able to track the state even
when the stability changes and its transitions from stable to
unstable. 
\section{Homogeneous Field}
\label{sec:homogeneous}

For a square system in a homogeneous field a systematic analysis of
the stable and unstable patterns has been performed in
\cite{SAV:2012:NBS}. The paper shows how the patterns change as the
strength of the applied field changed. Especially the transitions
between stable and unstable states have been studied in detail.

We reproduce in Figure~\ref{fig:small} the results for a small domain
with edge length $d=3\xi$.  This system is so small that it only
supports a few vortices. The figure shows the free energy as a
function of the strength of the homogeneous magnetic field.  The curve
$A$ starts at $\mu = 0$ where the system is in the homogeneous
solution $\psi\equiv 1$ and is completely superconducting. The
solution has all the symmetries of the system and is stable as a
global minimum of the free energy.

For nonzero field strength, the solutions deviate from the homogeneous
superconducting state, developing zones of lower supercurrents
near the edges of the domain.  
At field strength $\mu \approx 1.64$, an eigenvalue with
multiplicity 2 becomes unstable. 

At this bifurcation point two
different families of solution branches emerge. Both are unstable and
correspond to a single vortex moving towards the middle of the
domain. Along branch $J$ a single vortex moves towards the center
along one of the center lines. Along branch $K$ a single vortex moves
along one of the diagonals.

Both connect with branch $B$ in the other bifurcation point.  This
branch has one vortex in the middle of the domain but it is unstable
for small magnetic fields.  It becomes stable at the second
bifurcation point and it remains stable until the superconductivity is
lost for large field strengths.

The two curves $J$ and $K$ that connect the two bifurcation points
correspond to maxima in the free energy curve while the solid curves
correspond to minima. In that light the calculation of these unstable
states and their corresponding energy is valuable since it gives an
indication of the energy that is required to cross the barrier between
two stable states for a fixed magnetic field. Indeed, a transition
between the stable state without vortices and the stable state with a
vortex would require a transition where a vortex enters the domain.
The energy of this state is given in Fig.~\ref{fig:small}.

For larger systems in a homogeneous field similar bifurcation diagrams
can be constructed and there are many connections between the
bifurcation points.  We refer to \cite{SAV:2012:NBS} for a complete
discussion of the stable and unstable states of a square system in a
homogeneous field.

\section{Square domain under a disc with radius $2\xi$}
\begin{figure}
\begin{center} Curve $A$
\end{center}
\begin{tabular}{cccc}
  \includegraphics[width=0.12\textwidth]{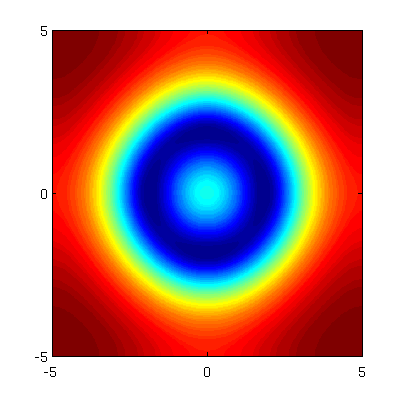}&  \includegraphics[width=0.12\textwidth]{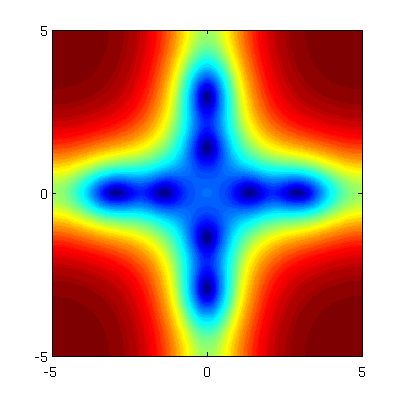}&  \includegraphics[width=0.12\textwidth]{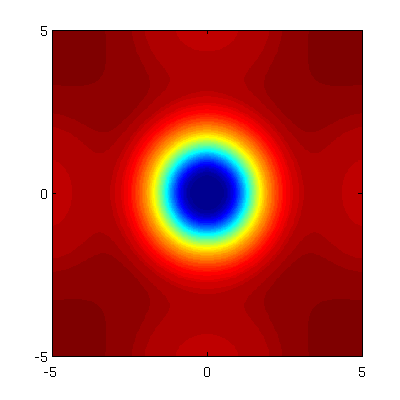}&  \includegraphics[width=0.12\textwidth]{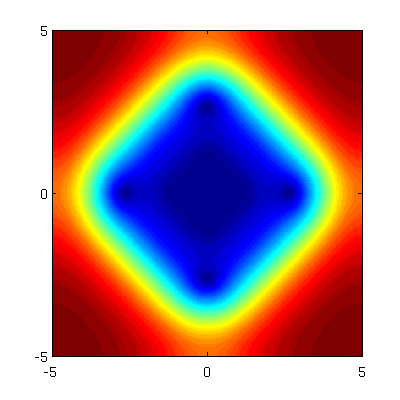}\\
    \includegraphics[width=0.12\textwidth]{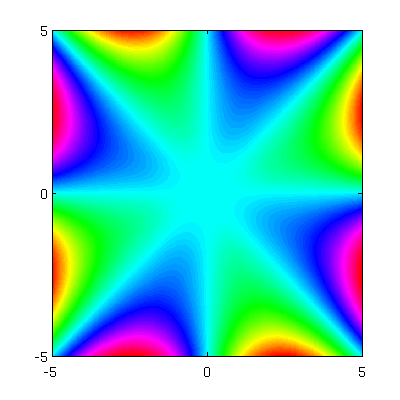}&  \includegraphics[width=0.12\textwidth]{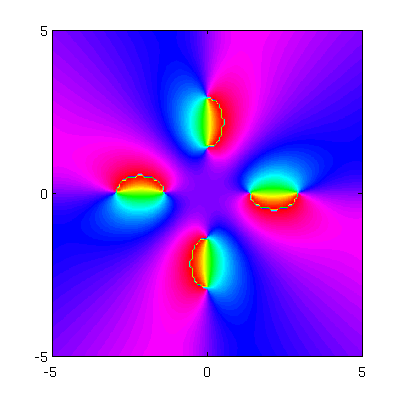}&  \includegraphics[width=0.12\textwidth]{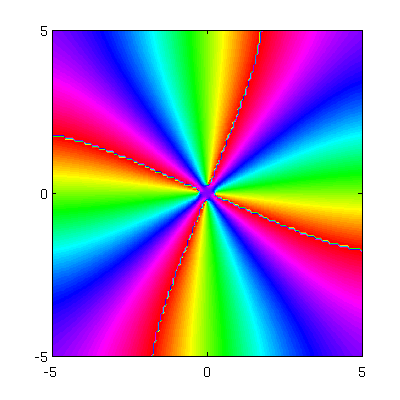}&  \includegraphics[width=0.12\textwidth]{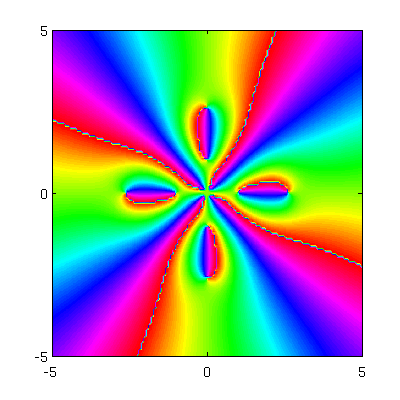}
\end{tabular}
\begin{center}Curve $B$
\end{center}
\begin{tabular}{cccc}
  \includegraphics[width=0.12\textwidth]{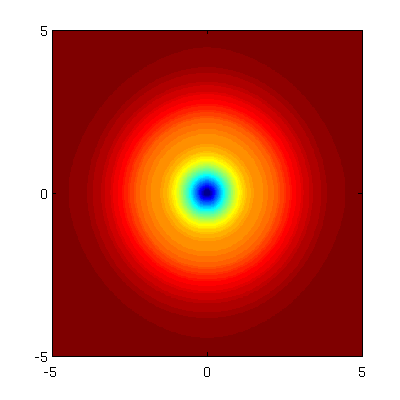}&  \includegraphics[width=0.12\textwidth]{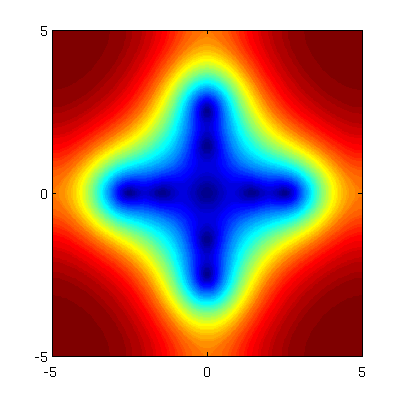}&  \includegraphics[width=0.12\textwidth]{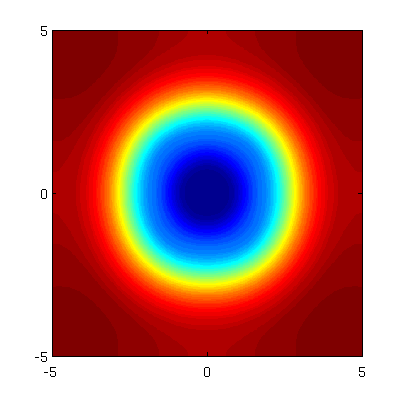}&  \includegraphics[width=0.12\textwidth]{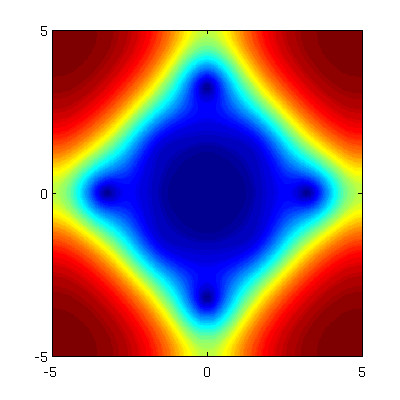}\\
    \includegraphics[width=0.12\textwidth]{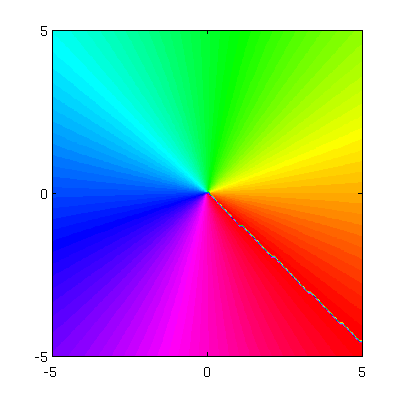}&  \includegraphics[width=0.12\textwidth]{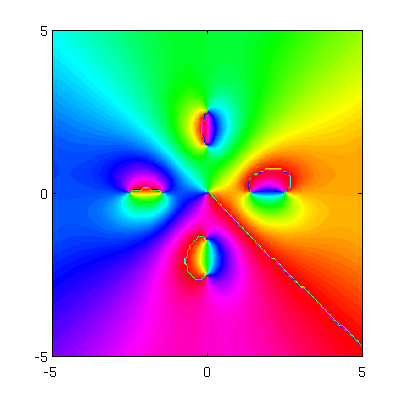}&  \includegraphics[width=0.12\textwidth]{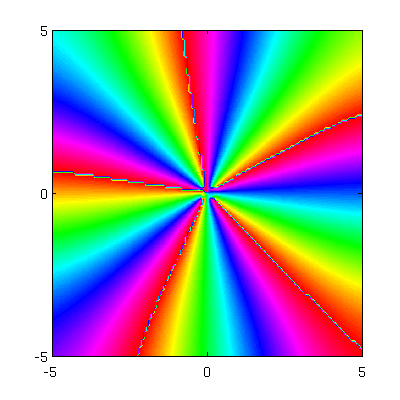}&  \includegraphics[width=0.12\textwidth]{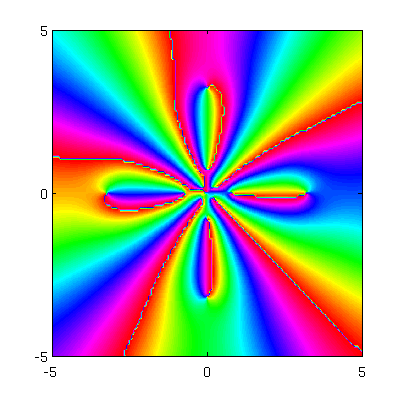}
\end{tabular}
\begin{center}Curve $C$
\end{center}
\begin{tabular}{cccc}
  \includegraphics[width=0.12\textwidth]{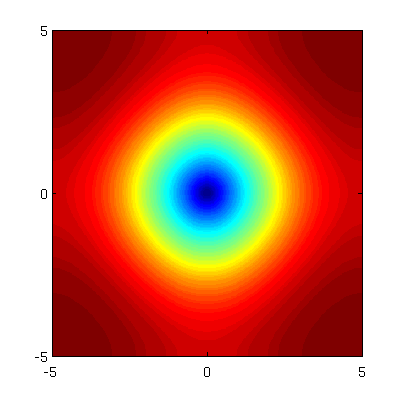}&  \includegraphics[width=0.12\textwidth]{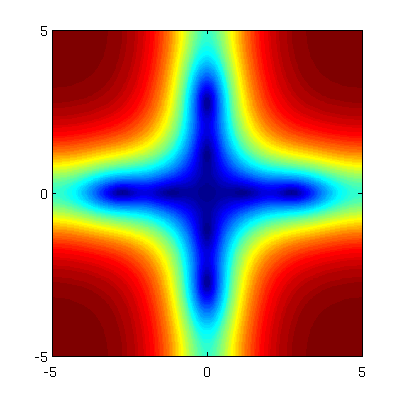}&  \includegraphics[width=0.12\textwidth]{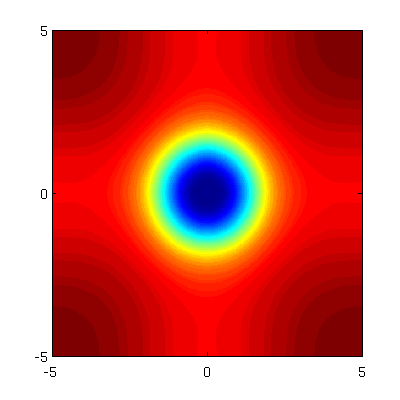}&  \includegraphics[width=0.12\textwidth]{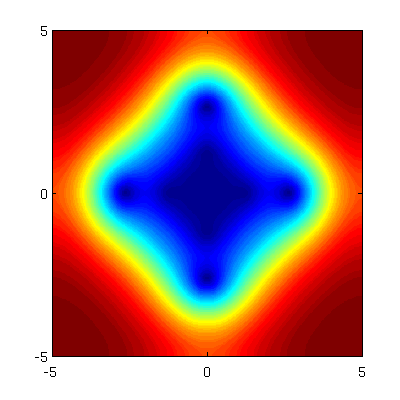}\\
    \includegraphics[width=0.12\textwidth]{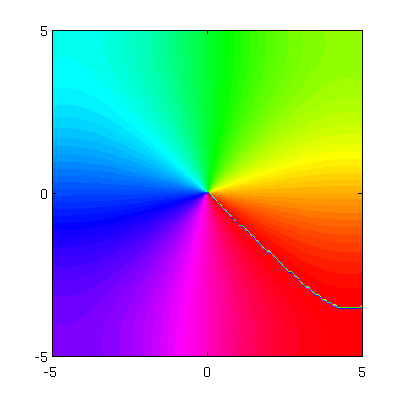}&  \includegraphics[width=0.12\textwidth]{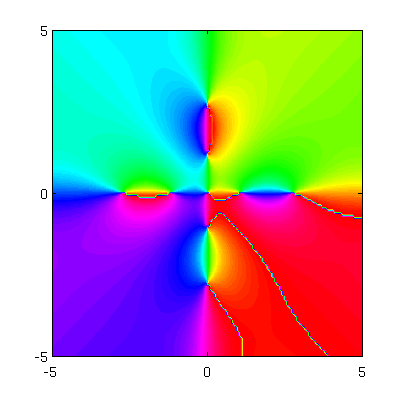}&  \includegraphics[width=0.12\textwidth]{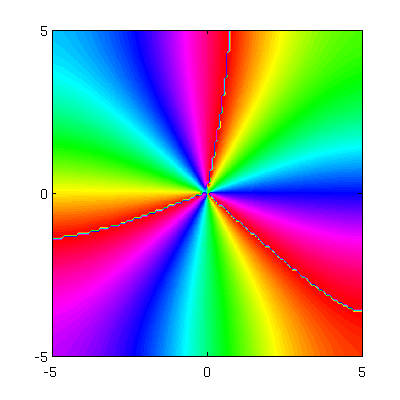}&  \includegraphics[width=0.12\textwidth]{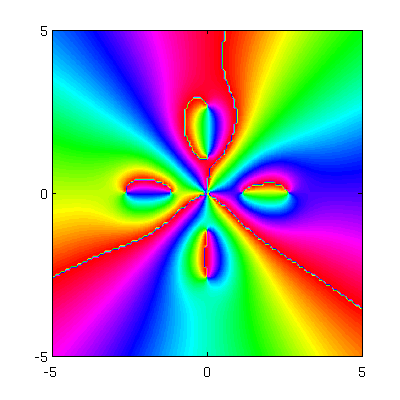}
\end{tabular}
\begin{center}Curve $D$
\end{center}
\begin{tabular}{cccc}
  \includegraphics[width=0.12\textwidth]{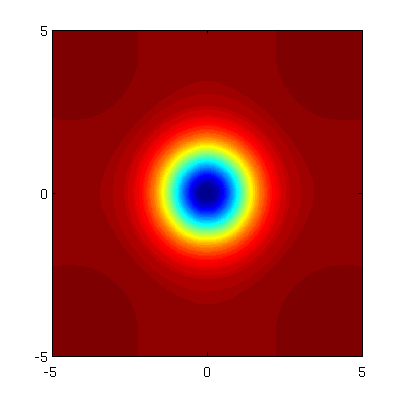}&  \includegraphics[width=0.12\textwidth]{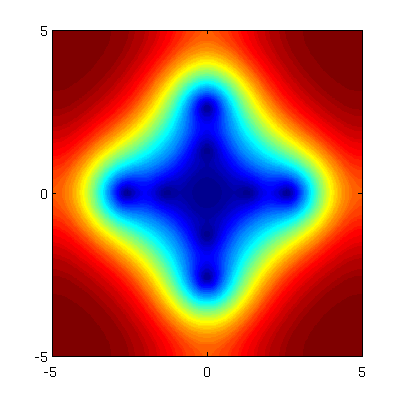}&  \includegraphics[width=0.12\textwidth]{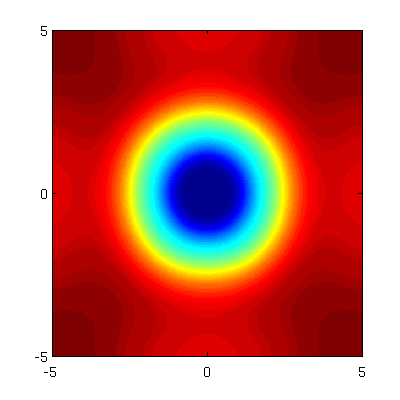}&  \includegraphics[width=0.12\textwidth]{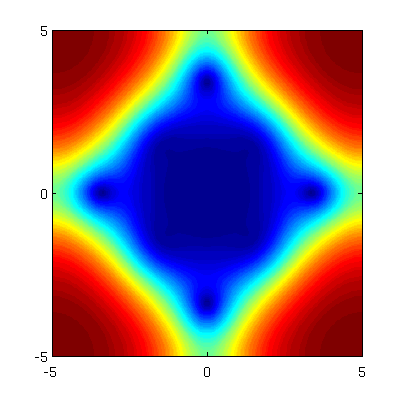}\\
    \includegraphics[width=0.12\textwidth]{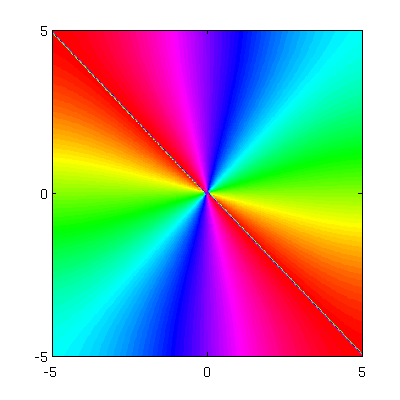}&  \includegraphics[width=0.12\textwidth]{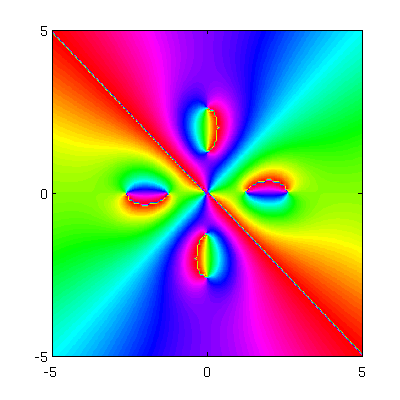}&  \includegraphics[width=0.12\textwidth]{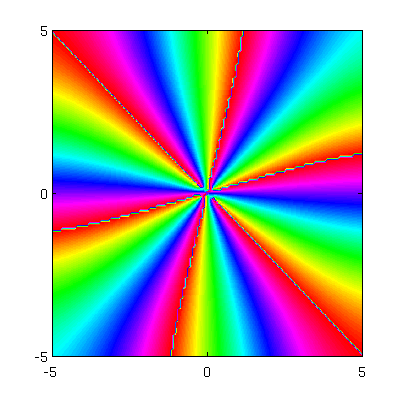}&  \includegraphics[width=0.12\textwidth]{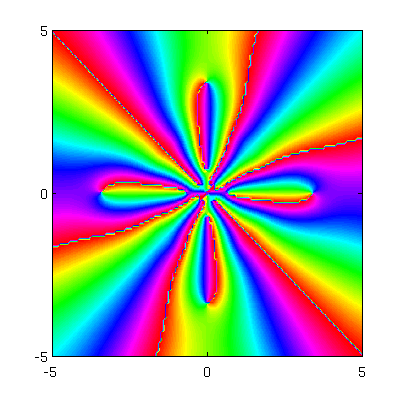}
\end{tabular}
\caption{(Color online) The patterns that are formed for a system with a magnetic disc
  of radius $r=2\xi$. We show the representative patterns along curve
  $A$, $B$, $C$ and $D$ for increasing magnetic field strength.  Each
  of the curves go through the same sequence. Under the edge of the
  magnetic disc four pairs of vortex and anti-vortex are formed. As the
  field increases further the four anti-vortices move out of the
  system and the four remaining vortices merge with the vortex that is
  already present in the middle to form a giant vortex.  Along this
  sequence the multiplicity is increased with four.  Then as the field
  increases further additional pairs of vortices and anti-vertices
  merge.}
\label{fig:small_dot}
\end{figure}

\label{sec:radius2.0}
In this section we look at a thin square superconducting domain with
size $10\xi \times 10\xi$. It sits under a superconducting disc of of
height $1\xi$ and radius of $2\xi$.  It sits a distance $0.1\xi$ above
the superconducting sample.

In Fig.~\ref{fig:rad2} we show the free energy of the states that
appear in the system as a function of the magnetic field. Initially
we discuss the states that have all the symmetries of the square
domain, i.e the symmetry group $D_4$ (see \cite{SAV:2012:NBS}).  There
are four curves with full symmetry denoted with labels $A$, $B$, $C$
and $D$.  The stability of the states along these curves changes in
bifurcation points denoted with $1$, $2$, $3$, $\ldots$.  The curves
$E$, $F$, $G$ and $H$ are connecting curves that connect bifurcation
points on different curves.

\subsubsection{Curve $A$}
Let us start with curve $A$. It starts at the trivial state $\psi
\equiv 1$ at zero magnetic field. As the field increases the free
energy of the curve rises and the solution starts to deviate from the
trivial solution.  Especially under the edge of the magnetic disc the
superconductivity is perturbed by the steep changes in the magnetic field.

The curve $A$ becomes unstable at $\mu=0.446$ in the bifurcation point
1. Here an eigenvalue of the Jacobian becomes positive real and the
system becomes unstable.  At this bifurcation point the curve $E$
connects to the branch to curve $B$. This saddle point curve is
unstable and has a single vortex solution. The connecting curves will
be discussed later in detail in Sec.~\ref{sec:connections}.

On curve $A$, next to bifurcation point 1, there is a second
bifurcation point 2 with a slightly increased strength of the magnetic
field where a second eigenvalue becomes positive real.  This second
bifurcation point connects through curve $F$, a branch that features
two vortices moving in, to branch $C$. Again, this curve will be
discussed later.

At fields around $\mu=0.591$ on curve $A$ the perturbations under the
edge of the magnetic disc become more pronounced and four
vortex-anti-vortex pairs are formed. These vortex-anti-vortex pairs form
where the center lines of the square intersect with the circle of the
edge of the magnetic disc. This is where the edge of the magnetic disc
is closest to the edges of the square domain.  The formation of these
pairs is shown in the top row of Fig.~\ref{fig:small_dot} where
representative patterns as shown along curve $A$ as the strength of
the magnetic field increases.

As the field increases further, the anti-vortex and the vortex in each
pair move apart. The anti-vortex moves along the center line of the
square away from the center and towards the edge of the domain and,
finally, it moves out of the sample.  During this sequence the free
energy curve goes through two turning points and the curve forms a
swallow tail as can be seen in Fig.~\ref{fig:rad2}.

At the same time, as the anti-vortices move out, the four remaining
vortices move towards the center of the domain where they merge into a
giant vortex of multiplicity four. Such a giant vortex is also shown
as one of the patterns for curve $A$ in Fig.~\ref{fig:small_dot}.

The curve $A$ regains stability at a field strength of $\mu=0.571$ in
bifurcation point 3.  The solution is now a giant vortex with
multiplicity 4.  And for the interval $0.895~\le~\mu~\le~1.175$ the
state has the smallest free energy of all states.

The curve remains stable up to field strengths $\mu=1.754$ where it
becomes unstable.  The curve again forms a swallow tail along which
four additional vortex anti-vortex pairs are formed while the giant
vortex with multiplicity 4 remains in the center of the system.  Again, the
anti-vortices move out and the vortices merge with the giant
vortex. It has a multiplicity 8.

So at each transition through a swallow tail the multiplicity of the
giant vortex is increased by four. Along the curve $A$ we can
identify regions of the magnetic field where a stable giant vortex is
formed with multiplicity either $4$, $8$, $12$ and so on.

\subsubsection{Curve $B$, $C$ and $D$}\label{sec:connections} The
curve $B$ goes through a similar series of transitions as curve
$A$. The curve $B$ has for small magnetic fields a single vortex in
the middle of the domain. The states have all the symmetries of the
system.  Similarly to the curve $A$, as the field increases, the free
energy curve forms swallow tails where four vortex-anti-vortex pairs
are created and split into four vortices that merge with the vortex
already present in the center of the domain to form a giant vortex of
multiplicity 5.  The four anti-vortices move out of the domain along
the center lines. This sequence is illustrated in
Fig.~\ref{fig:small_dot}.

So, along the curve $B$ we can identify regions of the magnetic
field  where a stable giant vortex is formed in the center of the
domain with a multiplicity of, subsequently, 1, 5, 9, and so on. Near
the bottoms of the free energy curves the states are stable, near the
swallow tails they are unstable.  At every bifurcation point, where
the state transitions from stable to unstable, there are other
connecting curves, similar as in curve $A$.

The curve $C$, similar to curve $A$ and $B$, has intervals in the
magnetic field strength where stable giant vortices with multiplicity
2, 6, 10, 14, \ldots are formed. These regions are separated by
swallow tails similar to those found along the
curves $A$ and $B$. Note that the curve $C$ is closely related to
curve $B$. Indeed, if we continue curve $B$ in the negative $\mu$
direction we would find the curve $C$ reflected over the $y$-axis.

Also curve $D$ forms the sequence with giant vortices of multiplicity
$3$, $7$, $11$, ...  and shows similar swallow tails as the curves
$A$, $B$ and $C$.

The subsequent minima of the free energy form a repeating sequence, as
can be seen in the Fig.~\ref{fig:rad2}.  First, the curve $A$ forms
the minimum. For $\mu$ between 0.25 and 0.45 curve $B$ has the minimal
energy.  Then curve $C$ and $D$ has the state with the minimal energy.
Then it is the turn again to curve $A$ to form the pattern
with the minimal energy.  


\subsection{Connecting saddle point curves}
\begin{figure}
\begin{center}Curve $E$
\end{center}
\begin{tabular}{cccc}
  \includegraphics[width=0.12\textwidth]{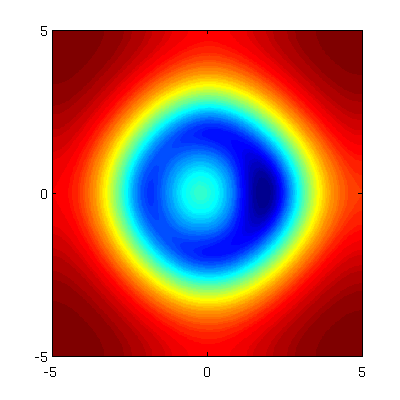}&  \includegraphics[width=0.12\textwidth]{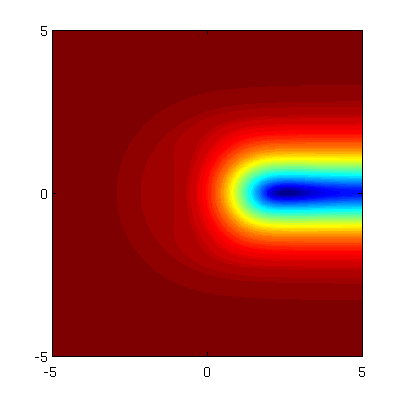}&  \includegraphics[width=0.12\textwidth]{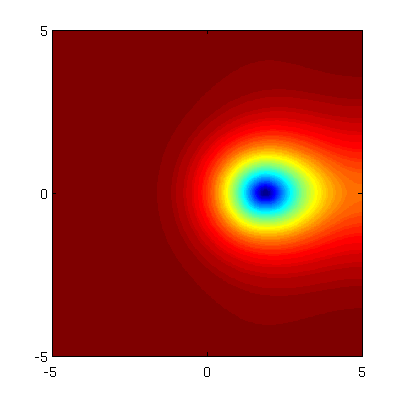}&  \includegraphics[width=0.12\textwidth]{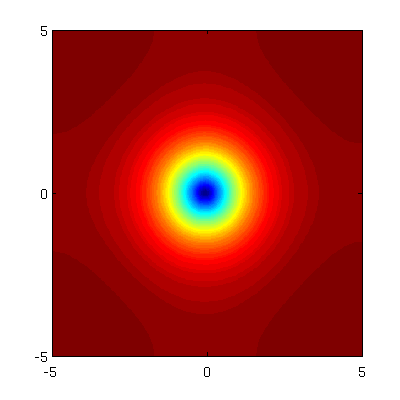}\\
    \includegraphics[width=0.12\textwidth]{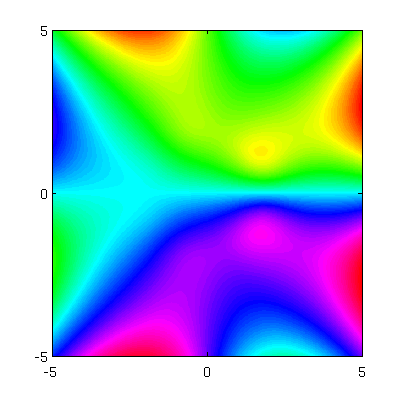}&  \includegraphics[width=0.12\textwidth]{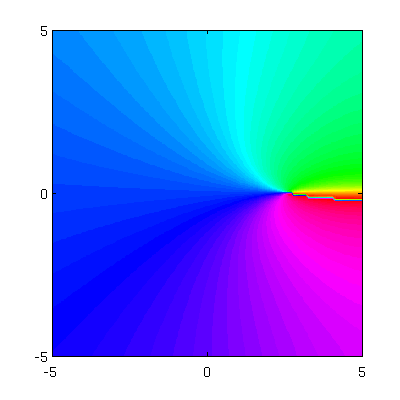}&  \includegraphics[width=0.12\textwidth]{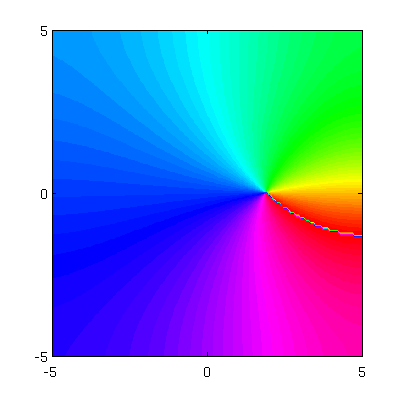}&  \includegraphics[width=0.12\textwidth]{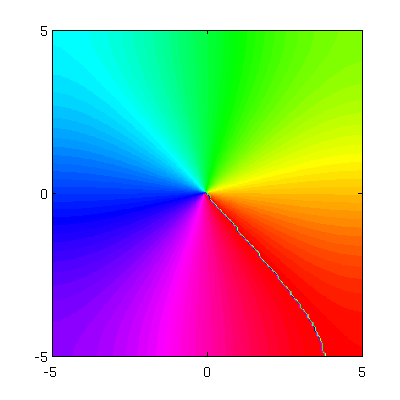}\\
\end{tabular}
\begin{center}Curve $F$
\end{center}
\begin{tabular}{cccc}
  \includegraphics[width=0.12\textwidth]{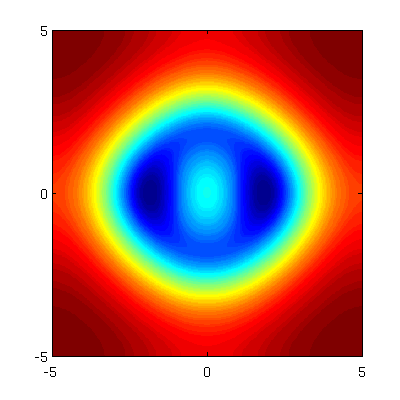}&  \includegraphics[width=0.12\textwidth]{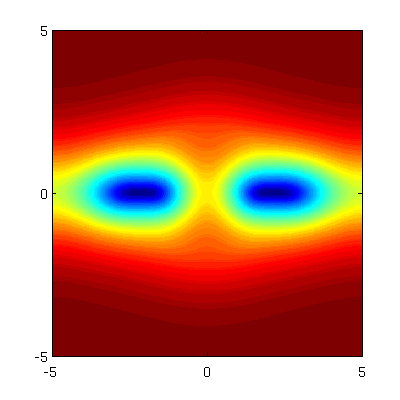}&  \includegraphics[width=0.12\textwidth]{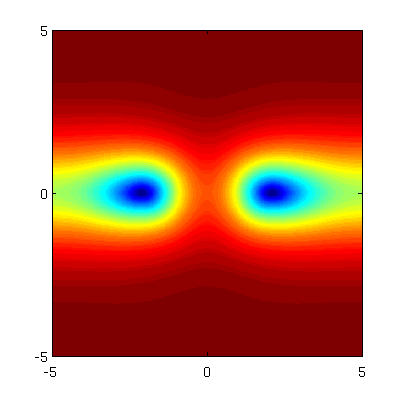}&  \includegraphics[width=0.12\textwidth]{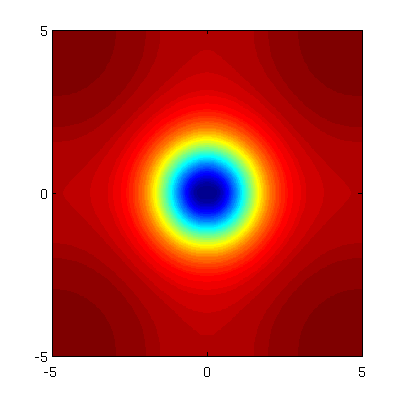}\\
    \includegraphics[width=0.12\textwidth]{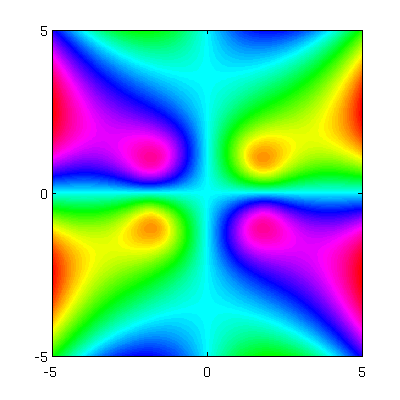}&  \includegraphics[width=0.12\textwidth]{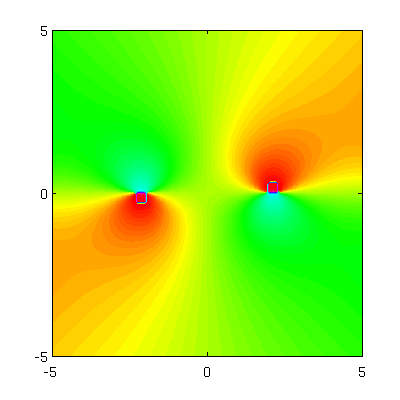}&  \includegraphics[width=0.12\textwidth]{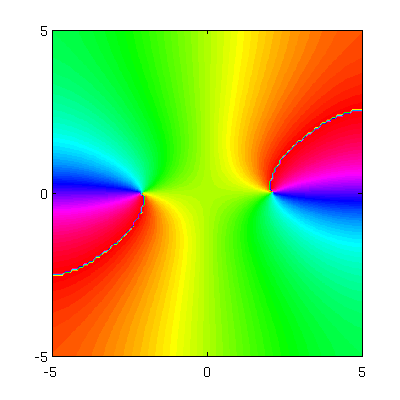}&  \includegraphics[width=0.12\textwidth]{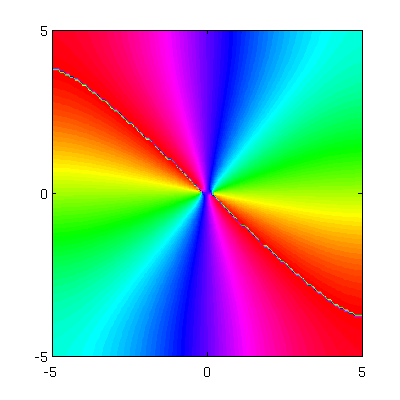}
\end{tabular}
\begin{center}Curve $G$
\end{center}
\begin{tabular}{cccc}
  \includegraphics[width=0.12\textwidth]{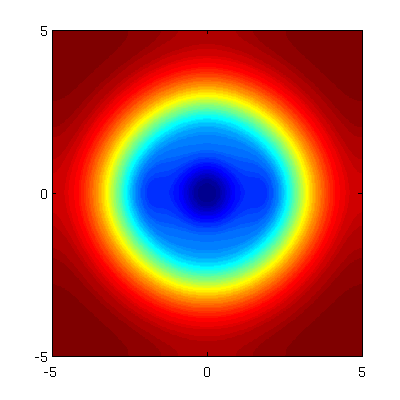}&  \includegraphics[width=0.12\textwidth]{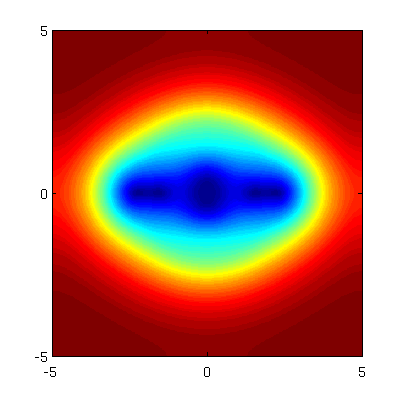}&  \includegraphics[width=0.12\textwidth]{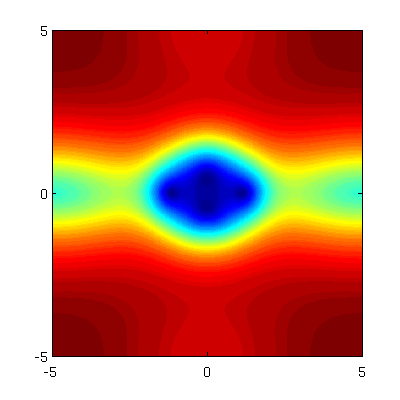}&  \includegraphics[width=0.12\textwidth]{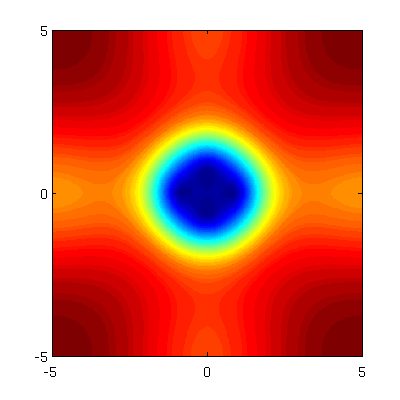}\\
    \includegraphics[width=0.12\textwidth]{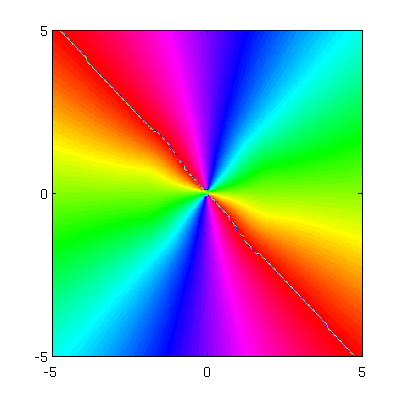}&  \includegraphics[width=0.12\textwidth]{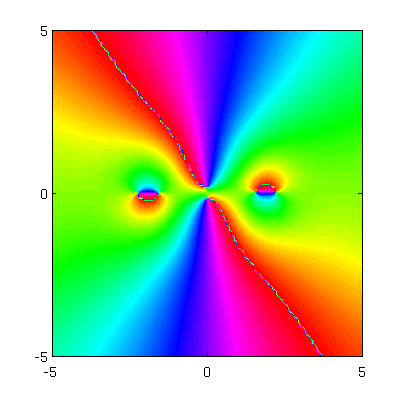}&  \includegraphics[width=0.12\textwidth]{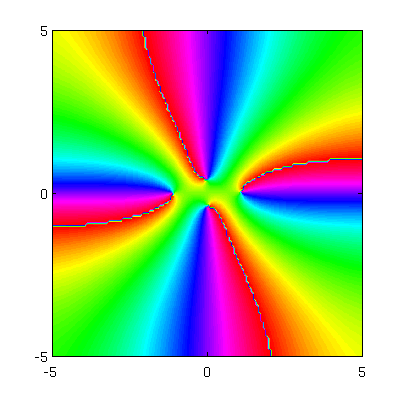}&  \includegraphics[width=0.12\textwidth]{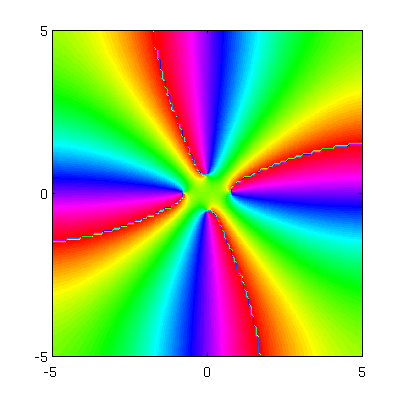}
\end{tabular}
\begin{center}Curve $H$
\end{center}
\begin{tabular}{cccc}
  \includegraphics[width=0.12\textwidth]{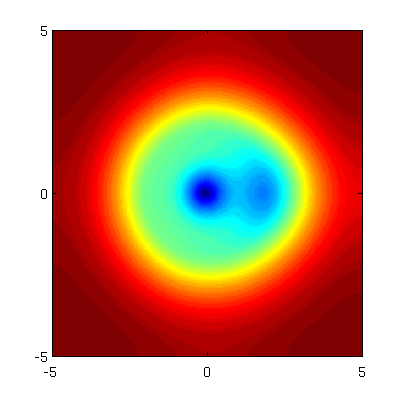}&  \includegraphics[width=0.12\textwidth]{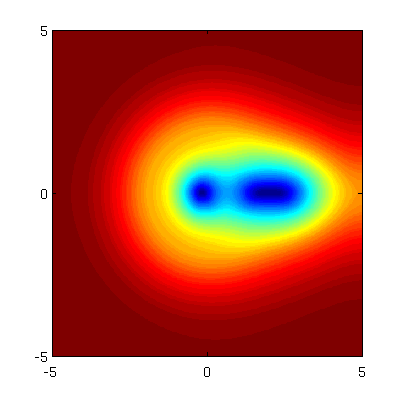}&  \includegraphics[width=0.12\textwidth]{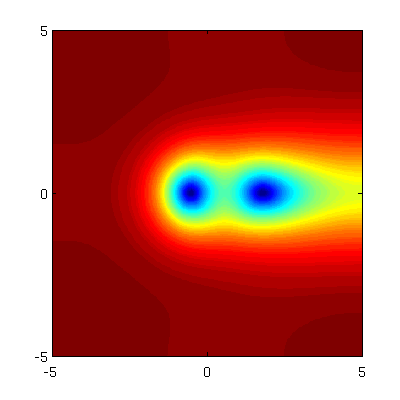}&  \includegraphics[width=0.12\textwidth]{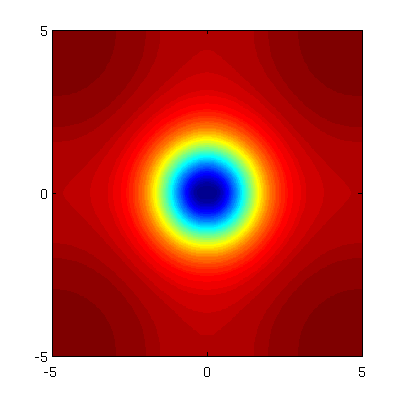}\\
    \includegraphics[width=0.12\textwidth]{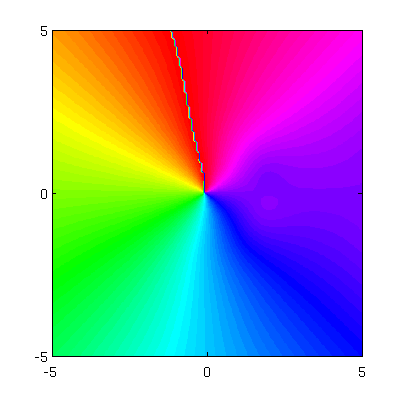}&  \includegraphics[width=0.12\textwidth]{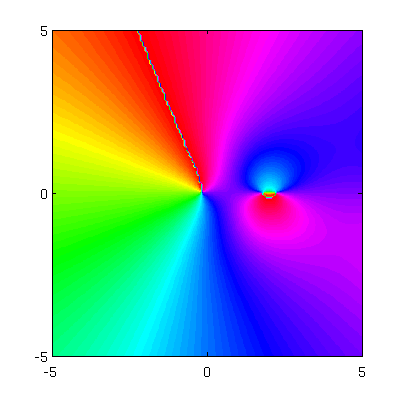}&  \includegraphics[width=0.12\textwidth]{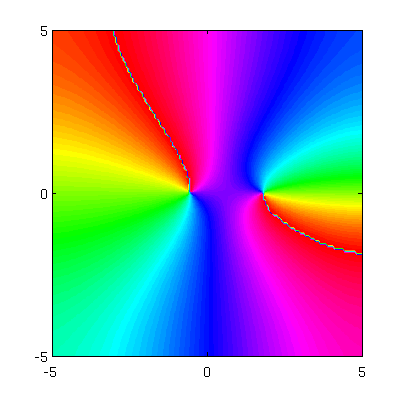}&  \includegraphics[width=0.12\textwidth]{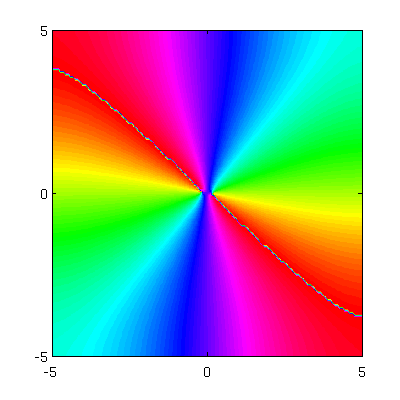}
\end{tabular}
\caption{(Color online) The patterns along the connecting curves $E$, $F$, $G$ and
  $H$ for a magnetic disc of radius $r=2\xi$.  These curve have a reduced
  symmetry compared to the patterns form Fig.~\ref{fig:small_dot}}
\label{fig:small_connecting}
\end{figure}

Along each of the curves there are several bifurcation points where
the states transition from stable to unstable.  At these bifurcation
points there is an eigenvalue of the Jacobian that crosses the origin
and becomes positive real. It is well known in bifurcation theory that
at each bifurcation point there is a connecting branch. For a complete
discussion on the bifurcations and the connections we refer 
to the literature on the equivariant bifurcation lemma and related results. 
A good introduction can be found in \cite{Hoyle:2006:PF}.

There are many bifurcations along the curves $A$, $B$, $C$ and $D$ and
we will not give a complete picture of all bifurcations. We only
focus on a few of the generic connecting curves that appear at the
first bifurcation points and the connections between
them.  Als these curves are unstable.

The curve $E$ connects the bifurcation point 1 on curve $A$ with
bifurcation point 6 on curve $B$, a curve with a single vortex in the
middle.  In bifurcation point 1 on curve $A$, there are no vortices in
the domain. but along the curve $E$ a single vortex-anti-vortex pair
is formed at one of the intersection points of the centerline and the
edge of the magnetic discs.  There are four possible positions of this
pair.  Then when the field is weakened the vortex and anti-vortex move
away from each other.  The anti-vortex leaves the domain and the
remaining vortex moves to the center of the domain where it connects
with the bifurcation point 6 on curve $B$.  A sequence of patterns
along this curve is shown in Fig.~\ref{fig:small_connecting}.

In a similar way there is a curve $F$ that connects the bifurcation
point 2 on curve $A$ with the bifurcation point 4 on curve $C$.  Along
this curve two pairs of vortex/anti-vortex are formed.  The two
anti-vortices move out and the two vortices merge into a giant vortex
in the middle of the domain at point 4. 

These same saddle point curves also appear between other curves.  For
example along curve $G$, which connects $B$ with $C$, a single vortex
moves in and merges with the already present vortex in to a giant
vortex with multiplicity 2 close to bifurcation 4. This is the
equivalent branch as curve $E$ that connects $A$ and $B$. 

Similarly, along curve $H$ that connects $C$ with $A$ two vortices move
to the center where they merge into a giant vortex with multiplicity 4
at bifurcation point 3 on curve $A$.

Again the free energy information about the unstable connecting saddle
node curves is useful for the understand the dynamics of a
superconductor. Because it gives an indication of the energies
required to make a transition between stable state state for a fixed
magnetic field.

%
%
\section{Square domain under a disc with radius $4.5\xi$}
\label{sec:radius4.5}
\begin{figure}
\begin{center}Curve $A$
\end{center}
\begin{tabular}{cccc}
\includegraphics[width=0.17\textwidth]{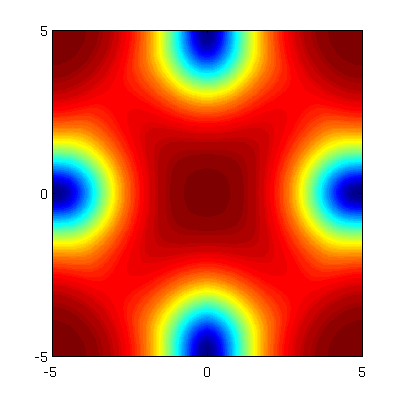}&  \includegraphics[width=0.17\textwidth]{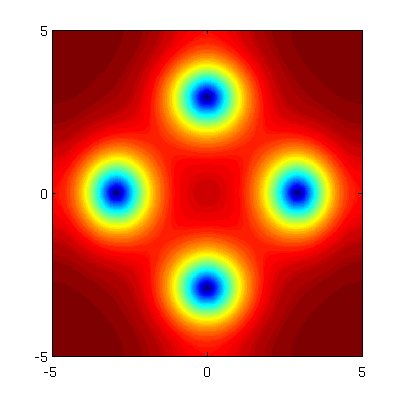}&  \includegraphics[width=0.17\textwidth]{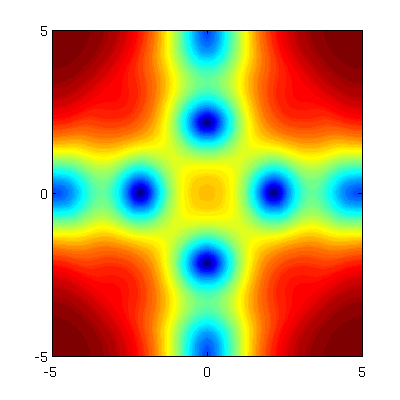}\\
\includegraphics[width=0.17\textwidth]{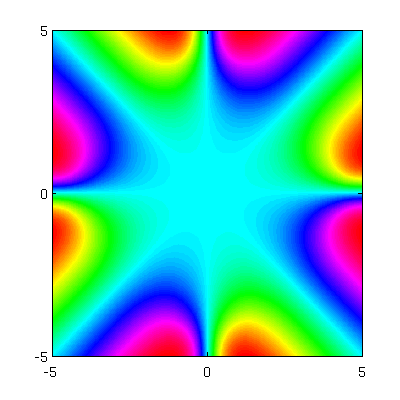}&  \includegraphics[width=0.17\textwidth]{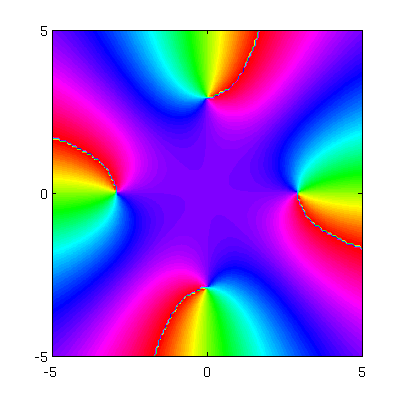}&  \includegraphics[width=0.17\textwidth]{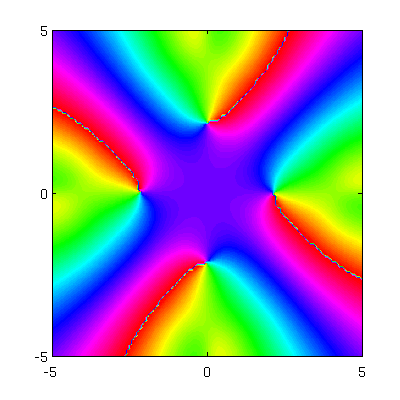}\\
\includegraphics[width=0.17\textwidth]{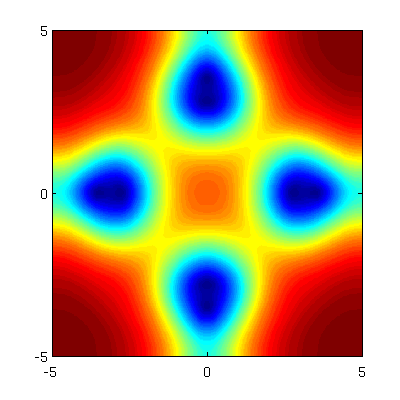}&  \includegraphics[width=0.17\textwidth]{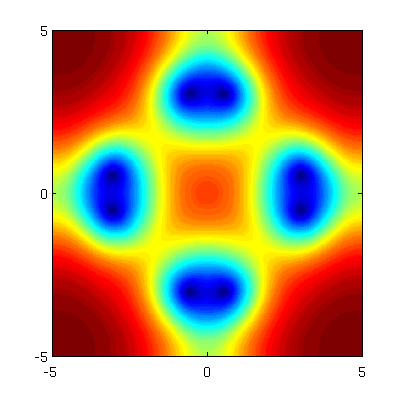}&  \includegraphics[width=0.17\textwidth]{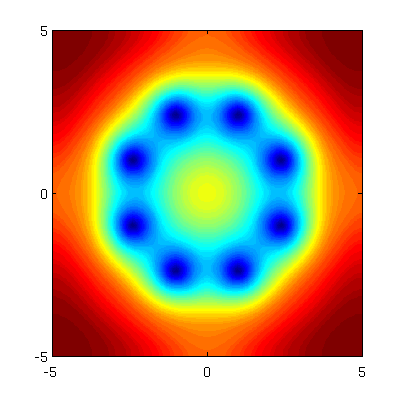}\\
\includegraphics[width=0.17\textwidth]{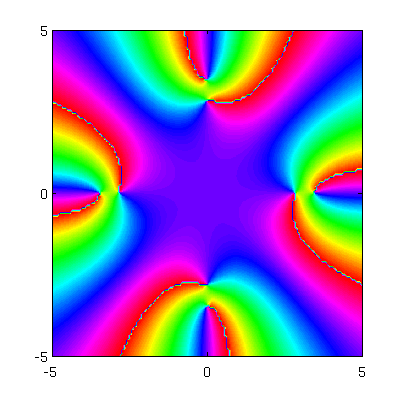}&  \includegraphics[width=0.17\textwidth]{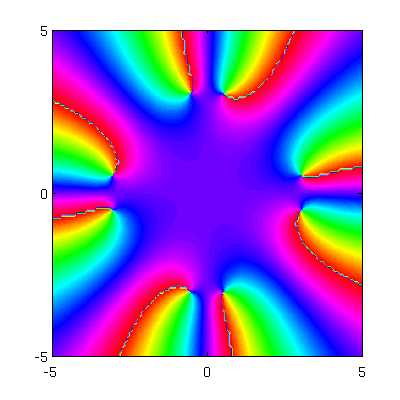}&  \includegraphics[width=0.17\textwidth]{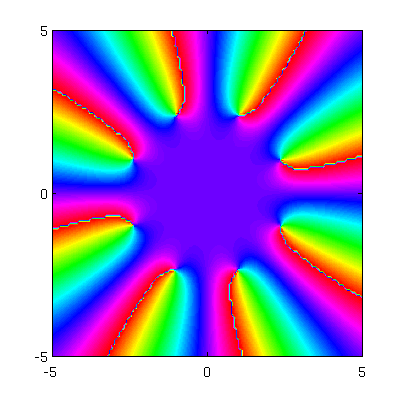}
\end{tabular}
\caption{(Color online) The patterns that appear along curve $A$ when the magnetic
  disc has a radius $r=4.5\xi$, almost equal to the size of the
  superconducting square. The main difference with the small magnetic
  disc is that there are no vortices formed in this system.  The field
  strength increases in lexicographic order from top left to bottom
  right.  First four vortices enter along the center lines of the
  square. However, they never merge into giant vortex in contrast to
  the small magnetic disc system. These four vortices form a stable
  pattern.  As the field increases further four additional vortices
  enter the system and they reorganize themselves along a ring.  This
  is also stable for a small range of magnetic fields.}
\label{fig:large_A}
\end{figure}
When the magnetic disc is larger, for example, close to the size of the
superconducting system, the vortices that appear underneath the
magnetic disc do not necessarily have to merge into a giant vortex with
a high multiplicity.  Indeed, there is enough space under the magnetic
disc to support multiple separated vortices.  As a result, the
transitions between stable and unstable states for a larger system
are different from thoses of the previous section.  In this section we study a
system with a radius of $r=4.5\xi$.

\begin{figure}
\includegraphics[width=0.5\textwidth]{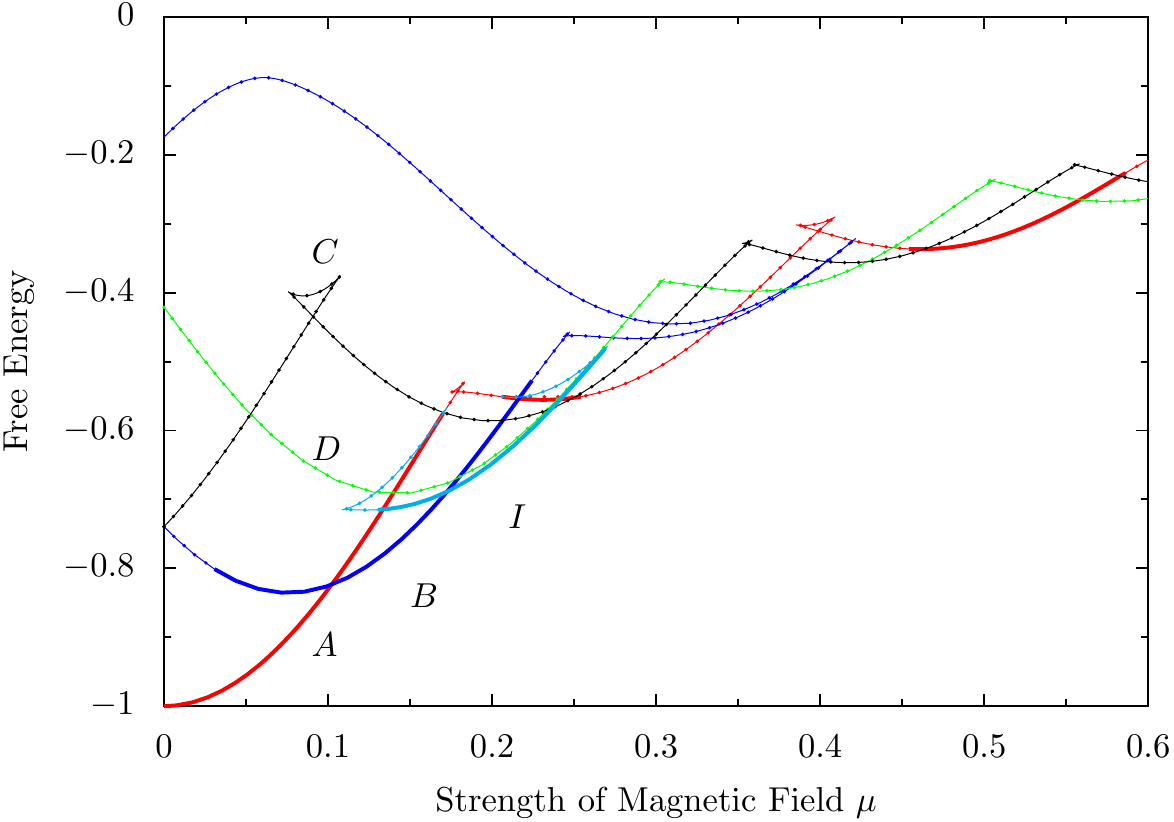}
\caption{ (Color online) Overview of the stable and unstable curves in a $10\xi\times
  10\xi$ system under a magnetic disc of radius $4.5\xi$. There are
  now much shorter stable regions in the curve $A$ and $B$, while the
  curves $C$ and $D$ are now completely unstable. Further more the
  connecting curves are now connecting bifurcation points in a
  different way }
\label{fig:large_overview}
\end{figure}

Fig.~\ref{fig:large_overview} gives an overview of the branches $A$,
$B$, $C$ and $D$ that were discussed earlier in
Sec.~\ref{sec:radius2.0} but are now shown for the larger disc. There
are, however, some important differences. Branch $C$ and $D$ are
completely unstable and never form a stable configuration of
vortices. This in contrast to the system with a small magnetic
disc. We will not discuss these patterns in detail.

\begin{figure}
\includegraphics[width=0.5\textwidth]{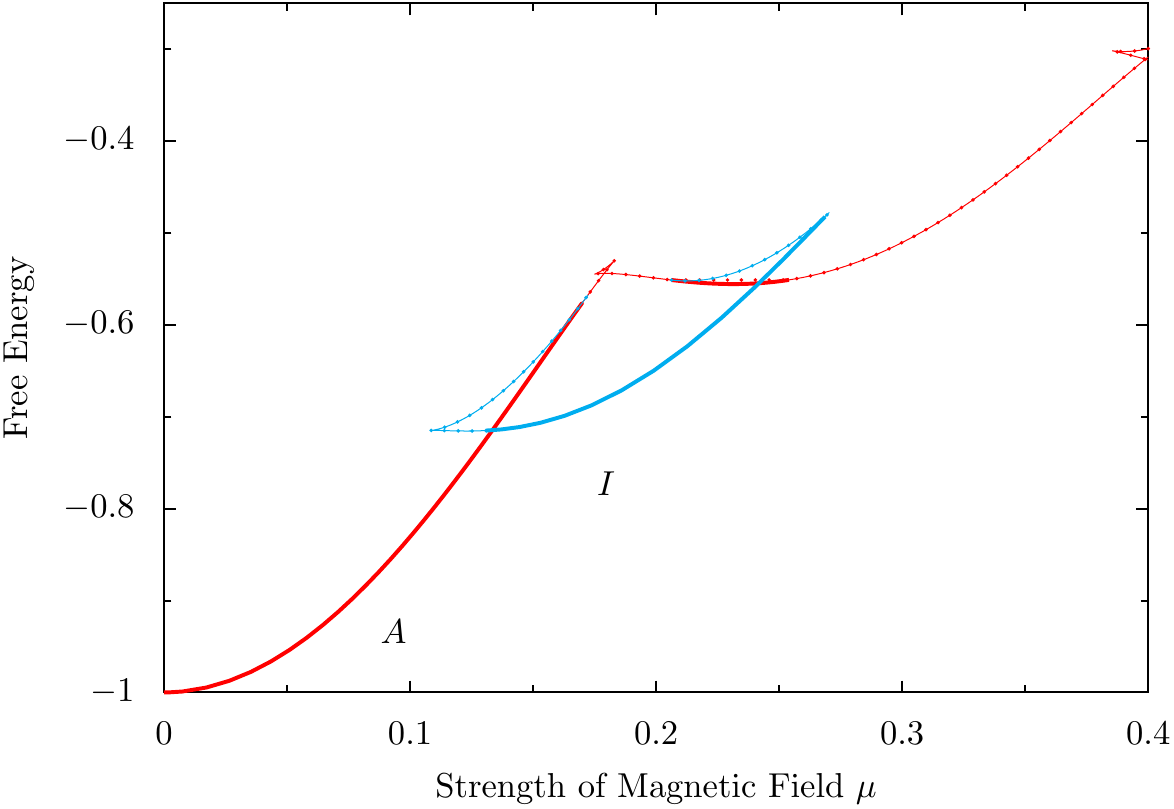}
\caption{ (color online) Detail of Fig.~\ref{fig:large_overview}. We show how curve
  $A$ loses and regains its stability at the the swallow tail. At the
  bifurcation points a connecting curve emerges that has a stable
  region. This in contrast to the results for the small system}
\label{fig:large_main}
\end{figure}

In Fig~\ref{fig:large_main} and Fig.~\ref{fig:large_A}, which shows the
corresponding patterns, we follow in detail the branch $A$ that starts
from the trivial solution $\psi \equiv 1$ at zero field. Also for this
system this state remains stable for small strengths of magnetic field.  At a bifurcation point 1 this state loses its stability.
Again there is another branch emerging from this bifurcation point.
However, now the branch does not connect to a bifurcation point on
another branch. Actually, the branch reconnects with the branch
$A$ in the bifurcation point $2$, where the branch $A$ regains
stability.

In Fig.~\ref{fig:large_A} we follow the pattern along $A$ for the
system with radius $4.5\xi$. In contrast to a small system it does not
support the formation of vortex anti-vortex pairs.  As the field
strengthens, four vortices move in along the center lines. This scenario is very similar
to what happens in a square superconducting system in a homogeneous
magnetic field \cite{SAV:2012:NBS}.  The four vortices form a stable
diamond configuration.  As the field strengthens further four
additional vortices move in and the eight vortices reorganize
themselves in a ring structure underneath the edges of the magnetic
disc.

\begin{figure}
\begin{center}Curve $I$
\end{center}
\begin{tabular}{cccc}
\includegraphics[width=0.17\textwidth]{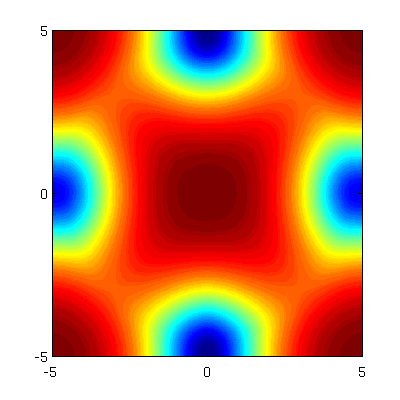}&  \includegraphics[width=0.17\textwidth]{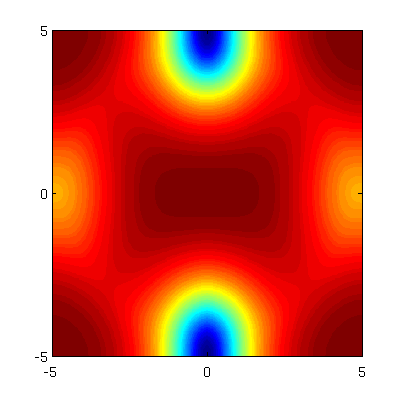}&  \includegraphics[width=0.17\textwidth]{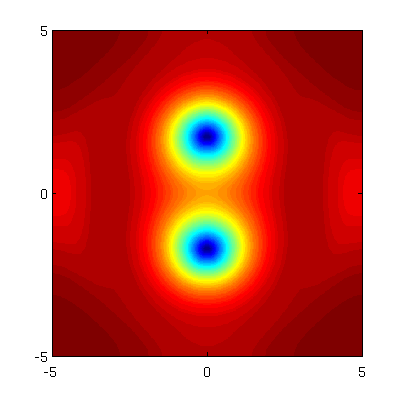}\\
\includegraphics[width=0.17\textwidth]{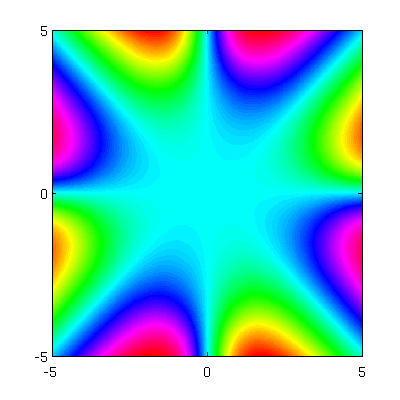}&  \includegraphics[width=0.17\textwidth]{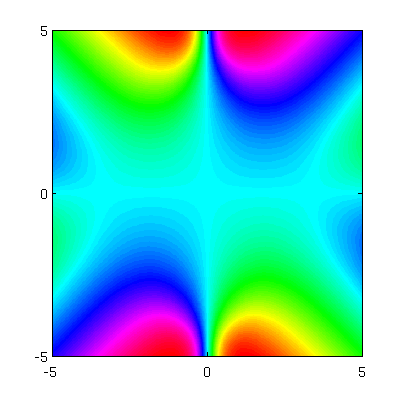}&  \includegraphics[width=0.17\textwidth]{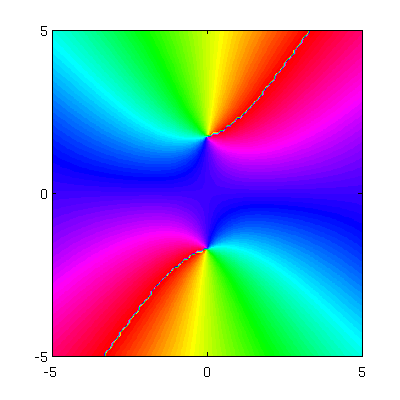}\\
\includegraphics[width=0.17\textwidth]{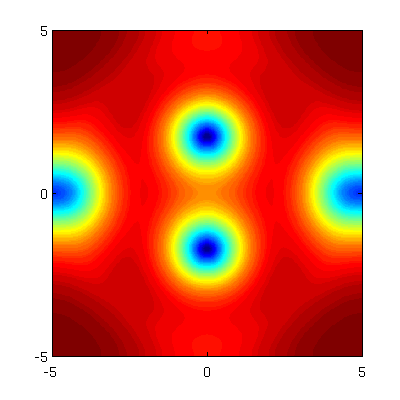}&  \includegraphics[width=0.17\textwidth]{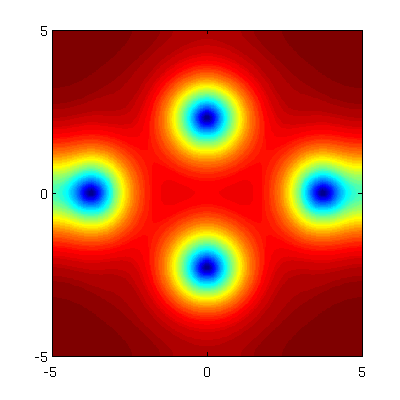}&  \includegraphics[width=0.17\textwidth]{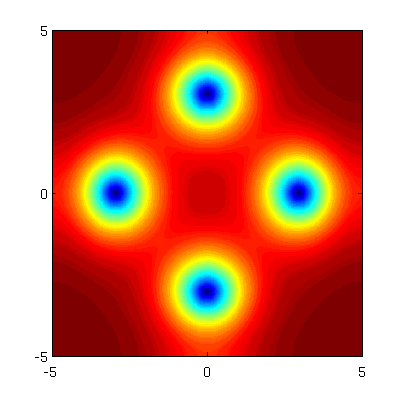}\\
\includegraphics[width=0.17\textwidth]{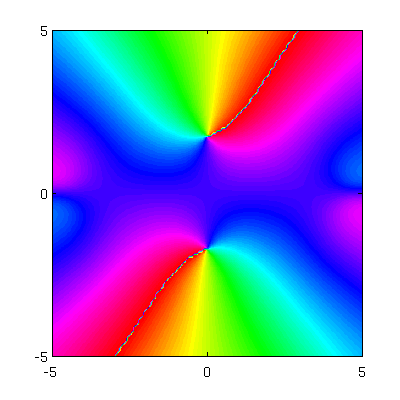}&  \includegraphics[width=0.17\textwidth]{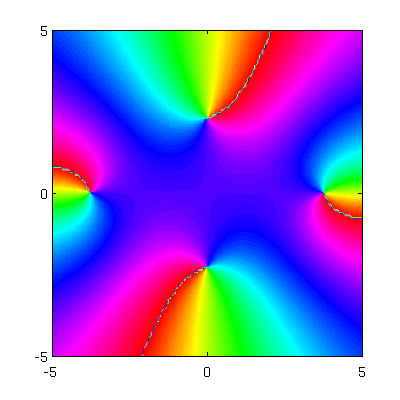}&  \includegraphics[width=0.17\textwidth]{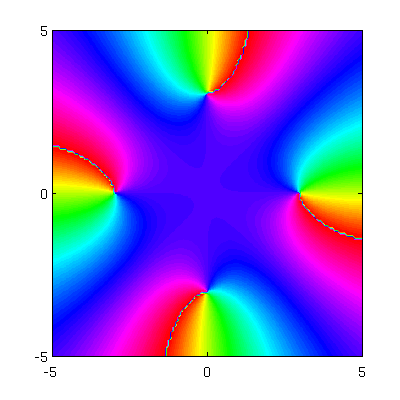}
\end{tabular}
\caption{(Color online) The patterns that appear along curve $I$ when the magnetic
  disc has a radius $r=4.5\xi$}
\label{fig:large_I}
\end{figure}

In Fig.~\ref{fig:large_I} we follow the branch $I$ that connects the
two bifurcation points on branch $A$ along this branch only two of the
four vortices enter the domain.  They move to the center as the field
increases then two additional vortices enter. The four vortices then
reorganize themselves into a diamond shape. At this point it reconnects 
with the curve $A$. 

\section{Discussion and Conclusions}
With the help of a preconditioned Newton-Krylov solver and numerical
continuation we have tracked the solutions of the Ginzburg-Landau
equations for a thin square extreme type-II superconductor in the
field of a magnetic dot. The method is able to track, as a function
of the magnetic field, both stable and unstable states. This enables
one to study the dynamics of the solutions of the Ginzburg-Landau
equations in an arbitrary geometry. The knowledge of the unstable
states further gives insight in the possible transitions between
stable states and the energy required to cross barriers that
separate them.

In this exploration of the free energy landscape we have identified
generic scenarios that appear repeatedly in the system. We have
studied a square superconductor of size 10$\xi\times$10$\xi$ under a
magnetic dot with radius $r=2\xi$ and radius $r=4.5\xi$. In the
first system, we have identified a generic scenario that repeats
itself over and over as the strength of the magnetic field
increases.  As a state is tracked as a function of the field
strength, the free energy forms a repeating sequence of swallow
tails.  Along such a swallow tail four pairs of vortex and
anti-vortices are formed. The anti-vortices move out of the domain
as the field strengthens and the four remaining vortices merge into
a giant vortex in the middle of the domain underneath the magnetic
dot. This generic scenario repeats itself for subsequent states
along their energy curves.

Furthermore, we have identified the saddle point states that connect
bifurcations points which appear on the main branches.  These
branches have reduced symmetry and have one or two vortices moving
in the domain. We have not identified all the possible connecting
curves, but we expect that there will be connecting saddle point
curves with three vortices moving in. Those have been already been
observed for samples in a homogeneous field \cite{SAV:2012:NBS}.

For the large magnetic dot, with radius $r=4.5\xi$ above scenarios
no longer hold because there is enough room underneath the magnetic
disc to form separate vortices rather than to collapse into a single
giant vortex. At the same time, since the size of the magnet is
almost the same as the superconducting domain, the field which the
superconductor experiences is practically homogeneous. As a result,
the stability properties of the main energy curves are completely
changed. Furthermore, the connecting saddle curves form different
connections, where appearance of antivortices is scarce.

The obvious continuation of the present work is the extension of the
study to 3D systems, based on the efficiency of the solver as shown
in Ref. \cite{schlomer2012efficient}. The current work lays the
foundation for a systematic bifurcation analysis of vortex rings and
loops that may appear around magnetic inclusions in 3D
superconductors, as studied in e.g. Ref. \cite{doria2007threefold}. The
plethora of there expected stable and unstable states guarantees a
promising exploration avenue to follow.

\acknowledgements We acknowledge support from FWO Vlaanderen through
project G.0174.08N.

\bibliography{refs}
\end{document}